# Negative Local Partial Density of States

Kanchan Meena[1,a], Souvik Ghosh[2,b], P Singha Deo[1,c]

[1] Department of Physics of Complex Systems, S. N. Bose National Centre for Basic Sciences, Kolkata, West Bengal, India.
[2] Department of Physics, Adamas University, Kolkata, West Bengal, India.

[a] 1996.kanchanmeena@gmail.com
[b] souvikghosh2012@gmail.com
[c] singhadeoprosenjit@gmail.com

**Abstract**

Real quantum systems can exhibit a local object called local partial density of states (LPDOS) that cannot be proved within the axiomatic approach of quantum mechanics. We demonstrate that real mesoscopic system that can exhibit Fano resonances will show this object and also very counterintuitively it can become negative, resulting in the enhancement of coherent currents.

Received 30 January 2025; First Review 13 February 2025; Accepted 07 March 2025

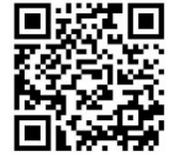

## Introduction

In recent times the subject of mesoscopic physics has attracted a lot of attention as it gives us an experimental route to probe some of the most fundamental problems in physics. As the name suggests, it is the regime that is in between the classical and quantum regimes. One such problem is that of defining a signal propagation time in quantum mechanics. This problem has been recently solved at theoretically making it completely consistent with the theory of relativity [1-5]. The puzzling feature that one has to accept as a result is that a signal can be also sent back in time, to the past. While this is counter-intuitive, a very convincing demonstration of it can be given and is presented in this work. If the signal propagation time between two points be divided by $\hbar$ then it gives the partial density of states (PDOS) which is a measure of the states through which the particle propagates between the two points. This can be shown from propagation of wave-packets in time [1, 3]. Therefore, if the PDOS become negative then the signal propagation time also becomes negative. We will demonstrate this quantity can be negative in this work which too is equally counterintuitive but consistent with all the theories of physics.

## The System

For the purpose of this demonstration, we consider the three-probe set up as described below with respect to Fig. 1.

The mesoscopic sample is the shaded region which is typically made of semi-conductor or metal, through which a current is passed. The leads $\gamma$ or $\alpha$ are current leads that may be made up of the same material. Lead $\beta$ is an STM tip that can attach to different points in the sample and is such that we can vary its proximity to the sample. The STM tip can thus be set up in many different ways but we will focus on one particular case. That is when the STM tip is weakly coupled to the sample at a point $r$ through a tunnelling barrier.

Let chemical potential of left reservoir connected to lead $\gamma$ be $\mu_\gamma$, chemical potential of right reservoir connected to lead $\alpha$ is $\mu_\alpha = 0$, and the other end of the the lead $\beta$ is also earthed. For incident energy $E$ such that $0 = \mu_\alpha < E < \mu_\gamma$, only lead $\gamma$ injects a current according to our classical notions. For a mesoscopic sample, this defines a quantum mechanical scattering problem. Such a set up can be obtained in the laboratory and we want to address the coherent current flowing from $\gamma$ to $\alpha$. It was shown [6] that in such a situation

$$|s'_{\alpha\gamma}|^2 - |s_{\alpha\gamma}|^2 = -4\pi^2 |t|^2 \vartheta_\beta \rho_{lpd}(E, \alpha, r, \gamma) \quad (1)$$

where

$$\rho_{lpd}(E, \alpha, r, \gamma) = \frac{-1}{2\pi} |s'_{\alpha\gamma}|^2 \frac{\delta \theta_{s'_{\alpha\gamma}}}{\delta U(r)} \quad (2)$$

is known as the local partial density of states (LPDOS), where $U(r)$ is the electrostatic potential at the point $r$,

$$\theta_{s'_{\alpha\gamma}} = arctan\left[\frac{\Im[s'_{\alpha\gamma}]}{\Re[s'_{\alpha\gamma}]}\right] \quad (3)$$

and $\frac{\delta\theta_{s'_{\alpha\gamma}}}{\delta U(r)}$ is a functional derivative. The parameter $t$ controls coupling of states of the sample with the states at the STM tip for which $\nu_\beta$ is the relevant density of states (DOS). For a commercially available STM tip $t$ and $\nu_\beta$ are either specified by the manufacturer or to be ingeniously determined by the user. Thus, $s'_{\alpha\gamma}$ is the scattering matrix element for scattering from $\gamma$ to $\alpha$. When the STM tip is removed by $t \to 0$ then this scattering amplitude will be $s'_{\alpha\gamma} = s_{\alpha\gamma}$. Coherent currents are given by $\frac{e_0}{h}|s'_{\alpha\gamma}|^2$ and $\frac{e_0}{h}|s_{\alpha\gamma}|^2$.

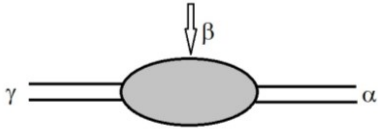

**Figure 1:** The basic setup for a three probe Landauer conductance where there are only two fixed leads indexed $\gamma$ and $\alpha$ apart from the STM tip $\beta$. This is a cartoon of a realistic experimental set up that help us measure some of the lower members in the hierarchy of DOS.

In earlier works, Eq. (1) could be justified in the semi-classical regime only which limits its value and applicability. The quantum regime remained unclear specially because in the quantum regime $\rho_{lpd}$, as calculated from Eq. (2), can become negative. We have taken the view that $\rho_{lpd}$ is a fundamental object that cannot be found within the axiomatic framework of quantum mechanics and in this work, we will show that it can be found in nature, in the so-called mesoscopic systems, in the quantum regime. Moreover, in the quantum regime of 1D, 2D and 3D, $\rho_{lpd}$ cannot be negative and its negativity is a speciality of the mesoscopic regime. Intuitively speaking, $\rho_{lpd}$ is positive and the RHS is negative accounting for loss of coherent electrons to the earthed lead $\beta$. Therefore, any change of sign in $\rho_{lpd}$ will result in a gain of coherent current due to the earthed STM tip. That can be regarded as supporting evidence for time travel because any intervention on a quantum state at an intermediate point, by an STM tip is supposed to only decohere the state. We will clarify as to why and when this can happen in real systems. Since LHS of Eq. (1) can be measured in the laboratory and so if the equality in Eq. (1) can be established then one can affirm that RHS is an objective reality that can be found in nature, although we cannot make sense of it within the axiomatic framework of quantum mechanics. The object $\rho_{lpd}$ cannot be defined within the axiomatic framework of quantum mechanics because it is refering to only those electrons that are coming from $\gamma$ and going to $\alpha$. The wave-function of an electron at a point $r$ can only be sensitive to either $\gamma$ or $\alpha$. But the object can be defined for a mesoscopic system using the idea of a physical clock. The idea of a physical clock as envisaged by Landauer and Buttiker [7] uses the idea that the spin of an electron can be thought of as a classical magnetic dipole and if it is subject to a classical force will presess like a classical dipole in a magnetic field. The classical force can be obtained from a potential of the form $\vec{\mu} \cdot B(\vec{r})$ rather than $\vec{\mu} \cdot \vec{B}(\hat{r})$ which essentially means the dot product is not the inner product of quantum mechanics but the usual dot product for a classical precession. The magnetic moment $\vec{\mu}$ is however taken to be quantized in units of $\frac{\hbar}{2}$. For a vanishingly small magnetic field, the expectation value of angular displacement can be calculated in a spinor space using the analyticity of scattering matrix elements. The angular displacement divided by the classical Larmor frequency of a dipole of strength $\frac{\hbar}{2}$ then defines a time for which a detailed derivation can be found in [5]. The equation of motion for the electron, that is the Schrodinger equation, is not used therefore, but only the analyticity of the spinor space is used. Thus, there is no direct way to verify this object from the wave-function but if it is averaged over all coordinates of the sample (that is averaged over $r$ and appropriate outgoing channels α then one arrives at an object called injectance of lead $\gamma$ which is completely in agreement with what one can calculate from the wave-function. There remains the doubt whether the non-averaged object as given in Eq. (2), at all has any relevance to reality.

Our work [1-5] shows that the basic premise for defining this object $\rho_{lpd}$ via a physical clock is that an open system is more general than a closed system. Meaning, it can only be defined if we have the leads $\gamma$ and $\alpha$ with their open ends connected to classical electron reservoirs. In that situation it can be shown that the object defined in Eq. (2) follow directly from the topology of the complex plane that is isomorphic to the Hilbert space (or the spinor space). The exact form of the term $\vec{\mu} \cdot \vec{B}(\hat{r})$ or the exact form of the dot product does not matter. It was also shown that negativity of $\rho_{lpd}$ is physical and corresponds to a wave-packet travelling back in time. Since a wave-packet can also carry information or signal, this raises some philosophical issues. Hence a more down to earth experimental verification will be of great help and that is what we will given in this work.

If we start from a closed system governed by Schrodinger equation, we cannot get an object that is $\gamma$ and $\alpha$ dependent because in that case the idea of a physical clock does not work as they inherently need asymptotic free states that are not a linear superposition of states. The classical reservoirs destroy the linear superposition of states in the leads while linear superposition exists inside the sample (shaded region). Events in the reservoir are all classical events and quantum event can only be defined in between two classical events in the reservoirs. This will be discussed again with respect to the 1D picture given in Fig. 2. Once we know $\rho_{lpd}$ we can use it to find other relevant quantities that we find in axiomatic framework of quantum mechanics and so we are not losing anything but only gaining some extra objects by this assertion. At any point we can go to the closed system by making the leads $\gamma$ and $\alpha$ very weakly connected to the system and recover all that we get by solving the Schrodinger equation for the closed system.

While the general view is that quantum mechanical laws are the fundamental laws of nature and classicality emerges, we adopt the view that a quantum event can only be defined between two classical events and the quantum laws can coexist with classical laws without any contradiction. This has some obvious but significant consequences and solves some long-standing problems in physics. First is that theory of relativity and quantum mechanics can co-exist consistently and there is no immediate need to quantize gravity. That is because now in quantum mechanics, like in theory of relativity, we can relate time intervals to signal propagation time. We can clearly obtain a coordinate time and a proper time within the framework of quantum mechanics. The points inside the classical reservoirs have local clocks giving their respective local times or coordinate time that is consistent with mass distribution in synchronizing the classical local clocks. A signal can propagate from one reservoir to another reservoir according to the laws of quantum mechanics and the time delay defines a proper time that can dilate and contract and can be different from local times clocked by the classical clocks inside the reservoirs. This is a consequence of the fact that a quantum event can only be defined between two classical events. Secondly, $\rho_{lpd}$ is a local object and removes the problem of collapse of wave-function, making it redundant, because measurements are only defined in the classical world in the reservoirs, while the time evolution of observables in the region between the reservoirs are to be determined quantum mechanically. To understand this, let us see what $\rho_{lpd}$ means. At the point $r$ there are electrons in a linear superposition of states. One may say that all these electrons came from lead $\gamma$ (assume for the time being that $\rho_{lpd}$ is positive definite) but some may be going to lead $\alpha$ (we call them dancing cats), some may be going to lead $\beta$ (we call them sleeping cats) and some may be getting reflected back to lead $\gamma$ (we call them beaten cats). At the point $r$, we have a linear superposition of dancing, sleeping and beaten cats. But we can now say something about only the dancing cats. They spend a time $\frac{-h}{2\pi}\frac{\delta\theta_{s'_{\alpha\gamma}}}{\delta U(r)}$ at the point $r$. At zero temperature there are $|s'_{\alpha\gamma}|^2$ of these dancing cats at the point, we can average over only the dancing cats to define a local partial density of states given by $\rho_{lpd}(E,\alpha,r,\gamma) = \frac{-1}{2\pi}|s'_{\alpha\gamma}|^2 \frac{\delta\theta_{s'_{\alpha\gamma}}}{\delta U(r)}$. That means we can selectively average over the dancing cats at the point $r$ and determine physical effects due to the dancing cats alone. In formal quantum mechanics, we cannot do this selective averaging for the dancing cats alone that are in a linear superposition with the sleeping and beaten cats. Note that $|s'_{\alpha\gamma}|^2$ and $\theta_{s'_{\alpha\gamma}}$ are measured at the junction of the sink reservoir which is a classical system.

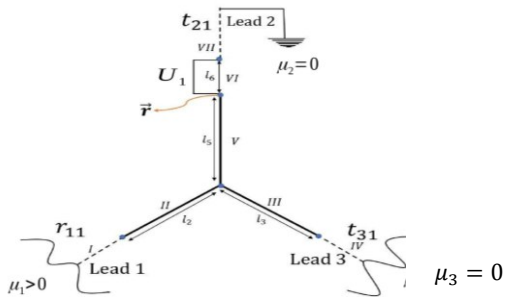

**Figure 2:** The sample is the three prong potential of shown by the solid lines and the entire system consist of the sample connected to three reservoirs via three leads. Different regions of the system is marked by Roman numbers, I, IV and VII being the leads, shown by dashed lines. Lead 2 is made exactly like that in Fig. 1 by earthing it and the chemical potential of lead 3 is also set to zero. In that case a small positive chemical potential $\mu_1$ will create a situation wherein only lead 1 inject a current while the two other leads carry some current away from the sample. Lead 2 connects to the sample through a tunnelling barrier shown as region VI. Lengths of different regions is shown as $l_2$ , $l_3$ etc. This system is a 1D version of the system in Fig. 1 where lead $\gamma$ is renamed as lead 1, etc, and as a result $t_{31} \equiv s_{31}$ , $t_{21} \equiv s_{21}$ , and $r_{11} \equiv s_{11}$ .

In this work we intend to justify Eq. (1) for the quantum regime, especially the situation when $\rho_{lpd}$ becomes negative. Note that there are two DOS that feature in Eq. (1), that are $v_\beta$ and an $r$ dependent DOS that is $\rho_{lpd}$. There is a factor $\frac{1}{2\pi}$ for each of them that cancels the factor $4\pi^2$. This is the factor $\frac{1}{2\pi}$ that appears in Eq. (2). There can be a very simple interpretation for this factor as follows. LHS of Eq. (2) is a DOS that can accommodate electrons that are countable but $2\pi\rho_{lpd}$ is uncountable, the RHS of Eq. (2) being in terms of $\theta_{s'_{\alpha\gamma}}(E)$ corresponding to a continuous rotation. Scattering probabilities $|s'_{\alpha\gamma}|^2$ and $|s_{\alpha\gamma}|^2$ are well defined in quantum mechanics and when multiplied by a factor $\frac{e_0}{h}$ they give the measured coherent current from $\gamma$ to $\alpha$. One of the facts that has emerged from the study of Fano resonances in [1-5] is that $\rho_{lpd}(E,\alpha,\boldsymbol{r},\gamma)$ can be positive as well as negative, implying that $|s'_{\alpha\gamma}|^2 - |s_{\alpha\gamma}|^2$ can be also positive as well as negative in the quantum regime. Hence, according to Eq. (1), when negative then the sample will lose coherent current to the earthed STM tip, and when positive it will gain coherent current from the earth which is counter-intuitive, as any invasive STM tip is supposed to lead to collapse of wave-function. As $|s'_{\alpha\gamma}|^2$ and $|s_{\alpha\gamma}|^2$ are both measurable and so if the equality in Eq. (1) can be justified, then in relative proportions $\rho_{lpd}(E,\alpha,\boldsymbol{r},\gamma)$ is also measurable which is good enough to confirm its sign. One may always try to justify Eq. (1) from a brute force practical experiment, but we will show below that it can be justified by what we call a theoretical experiment in the sense that Eq. (1) can be established from the topology of the complex plane formed by the imaginary and real parts of a scattering matrix element and hence independent of the quantum mechanical equation of motion as well as vague ideas of spin precession.

## The Model

For our theoretical experiment let us consider the system shown in Fig. 2. First of all, it is a system for which exact quantum mechanical calculations can be made and it is a system that exhibit pronounced Fano resonances. In this sense the system in Fig. 1 at Fano resonances with the modification that the upper lead is weakly coupled to earth via a strong potential can be represented by the 1D system in Fig. 2. Since it is a question of whether LPDOS is present at all in a quantum system and more importantly whether its negativity is allowed in quantum mechanics, then demonstrations in 1D quantum mechanics should suffice.

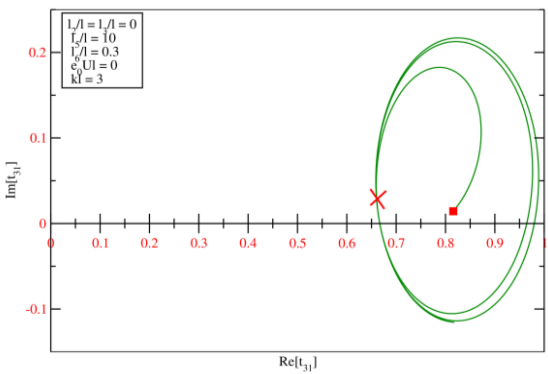

**Figure 3:** In this figure we plot the AD for $t_{31}$ of the system shown in Fig. 2 for the coupling potential $U_1$ varying in a range that give three sub-loops, all within one Riemann surface. The starting point is marked by a small square block corresponds to $U_1 l = -10$. The end point is unmarked and corresponds to a value $U_1 l = -1000$. All the sub-loops smoothly come back to a point marked by a cross. Other parameters are mentioned inside the figure.

Let us drop the factor $|t|^2 \nu_\beta$ from Eq. (1) because such factors will be only needed if we are using a commercially available STM tip, where this factor characterizes the STM tip. This requires us to drop a $2\pi$ factor in Eq. (1) for reasons mentioned in the previous paragraph. Which means for a first principle calculation for the system in Fig. 2 we expect (if the equality in Eq. (1) is correct) that

$$|s'_{\alpha\gamma}|^2 - |s_{\alpha\gamma}|^2 = -2\pi\rho_{lpd}(E,\alpha,\boldsymbol{r},\gamma) \quad (4)$$

Now to make it correspond to Fig. 2

$$|t'_{31}|^2 - |t_{31}|^2 = -2\pi\rho_{lpd}(E,3,\boldsymbol{r},1) \quad (5)$$

In Fig. 3, we plot the Argand diagram (AD) for $t_{31}$ for the system shown in Fig. 2 by varying the potential $U_1$ which connects the three-prong potential with lead 2 that is earthed, as $U_1$ is varied from $e_0 U_1 l = -10$ to $e_0 U_1 l = -1000$. We plot $\Im[t_{31}]$ versus $\Re[t_{31}]$, which is known as AD. The AD make smooth sub-loops within one Riemann surface and each sub-loop is due to a Fano resonance. Any physical quantity that depends on $t_{31}$ will go through a cycle over one particular sub-loop. For one closed sub-loop generated by monotonously varying a parameter, say $U_1(\boldsymbol{r})$, the local potential at $\boldsymbol{r}$,

$$\oint_c \Delta\,\theta_{t_{31}} = \oint_c \frac{\delta\theta_{t_{31}}}{\delta U_1(\boldsymbol{r})} \Delta U_1(\boldsymbol{r}) = 0$$

$$\text{where, } \theta_{t_{31}} = \arctan\frac{\Im[t_{31}]}{\Re[t_{31}]}$$

Or

$$-\oint_c \frac{1}{2\pi}|t_{31}|^2 \frac{\delta\theta_{t_{31}}}{\delta U_1(\boldsymbol{r})} \Delta U_1(\boldsymbol{r}) = \oint_c \rho_{lpd}(E,3,\boldsymbol{r},1)\Delta U_1(\boldsymbol{r}) = 0 \quad (6)$$

Similarly, over the same sub-loop,

$$\oint_c \frac{\delta|t_{31}|^2}{\delta U_1(\boldsymbol{r})}\Delta U_1(\boldsymbol{r}) = \oint_c \Delta|t_{31}|^2 = \oint_c \Delta|\Re[t_{31}] + \Im[t_{31}]|^2 = 0 \quad (7)$$

Therefore, from Eqs. (6) and (7) we get the equality

$$\oint_c \Delta|t_{31}|^2 = -\oint_c \frac{1}{2\pi}|t_{31}|^2 \frac{\delta\theta_{t_{31}}}{\delta U_1(\boldsymbol{r})}\Delta U_1(\boldsymbol{r}) = 0 \quad (8)$$

purely as a consequence of the topology of a complex plane $(\Im[t_{31}], \Re[t_{31}])$. If the Argand diagram trajectory would have enclosed the singularity, then the RHS of Eq. (6) would have been $2\pi$ instead of zero. Whatever be the terms in the Hamiltonian or the equation of motion, it is the topology of the complex plane that determines the outcome of the integration in Eq. 6. For the kind of Argand diagram in Fig. 3, that are smoothly closed within one Riemann surface, the equality in Eq. 8 is for the integrals. The integrands can well be completely different in magnitudes but the integrands will go through a positive-negative cycle over a sub-loop. Therefore, it is obvious that $\Delta|t_{31}|^2$ as well as $\rho_{lpd}$ will go through a positive-negative cycle over a closed sub-loop as a consequence of the topology of the relevant complex plane. Given the fact that LHS is physical, the cycles of $\rho_{lpd}$ is also physical as they are just different expressions of the same sub-loop made by the AD of $t_{31}$. If we want to compare the integrands in Eq. (8), then we need an extra factor of $2\pi$ as discussed before Eqs. (4) and (5). Thus,

$$\Delta|t_{31}|^2 \approx -2\pi \frac{1}{2\pi} |t_{31}|^2 \frac{\delta\theta_{t_{31}}}{\delta U_1(r)} \Delta U_1(r) \quad (9)$$

Viewing a derivative as an effect of an infinitesimal change in $U_1$ we get

$$|t'_{31}|^2 - |t_{31}|^2 \approx -|t'_{31}|^2 (\theta_{t_{31}} - \theta_{t'_{31}}) \quad (10)$$

where primed quantities and unprimed quantities are calculated for an infinitesimal difference of $U_1$. If we want to compare integrands then we can only write an approximate equality because we are comparing two different objects over one sub-loop. There will always be in the least a phase difference between the two quantities.

In Fig. 4, we plot LHS and RHS of Eq. (10) where the primed values and unprimed values are again for small differences in $U_1$ as the incident wave vector $k$ is varied and that is the horizontal axis. It becomes clear that LHS and RHS of Eq. (10) oscillate with positive and negative values. Initially the two curves are a little different but as $k$ increases, effects of dispersion is minimized, and magnitudes of the two curves are similar with a phase difference between them. The oscillations of both the curves are due to the smooth cyclic nature of the AD in Fig 3. In other words, the underlying principle behind the oscillation of both curves is the same. The two curves oscillate as a result of quantum interference. But the symmetric oscillation between positive and negative values and the similarity between the two curves can only be noted if there are well defined Fano resonances and signify the fact that the RHS of Eq. (1) is physical. For Breit-Wigner resonances the last two features are not seen.

## Conclusions

Thus, to answer the question why and when the equality of Eq. (1) can be observed, one can say that when we have closed sub-loops within one Riemann surface for the AD of the scattering matrix element of a relevant channel like $t_{31}$, then the equality will be found. Such closed subloops occur if there is a Fano resonance. When other type of resonances like the Breit-Wigner resonances start interfering with the Fano resonances then two things can happen. First is that the subloops are not smoothly coming back to the point depicted by a cross mark in Fig. 3 or we do not get closed subloops at all. In that case the equality in Eq. (1) will not be strictly valid but still there can be an approximate equality. Since the smooth subloop signify analyticity of a complex quantity like $t_{31}$, the equality of the two curves in Fig. 4 is not a direct consequence of quantum dynamics and thus we can prove the equality of Eq. (1), in the presence of Fano resonances without invoking the axioms of quantum mechanics. In this sense it is a theoretical experiment.

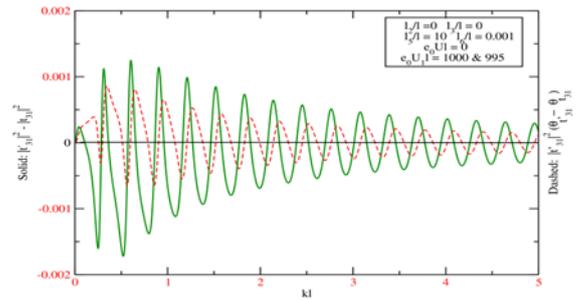

**Figure 4:** In this figure we are plotting the LHS and RHS of Eq. (10) to show that they both can oscillate between positive and negative values. The primed and unprimed values are for small differences in $U_1$ at a $k$ value that we vary continuously. The sign change and magnitude of both the curves originate from the smooth cyclic AD of Fig. 3.